\documentclass[%
 reprint,
 amsmath,amssymb,
 aps,
]{revtex4-2}

\usepackage{graphicx}
\usepackage{dcolumn}
\usepackage{bm}

\begin{document}

\preprint{APS/123-QED}

\title{Dynamical stabilization of Multiplet Supercurrents in Multi-terminal Josephson Junctions}

\author
{Ethan G. Arnault$^1$*, Sara Idris$^2$,Aeron McConnell$^2$, Lingfei Zhao$^1$,\\ Trevyn F.Q. Larson$^1$, Kenji Watanabe$^3$, Takashi Taniguchi$^3$,\\ Gleb Finkelstein$^1$, Fran\c{c}ois Amet$^2$*
\\
\normalsize{$^{1}$Department of Physics, Duke University, Durham, 27701, NC, USA}\\
\normalsize{$^2$Department of Physics and Astonomy, Appalachian State University, Boone, 28607, NC, USA}\\
\normalsize{$^{3}$Advanced Materials Laboratory, NIMS, Tsukuba, 305-0044, Japan}\\
\normalsize{$^\ast$To whom correspondence should be addressed; E-mail:  ega7@duke.edu, ametf@appstate.edu}
}

\date{\today}

\begin{abstract}
The dynamical properties of multi-terminal Josephson junctions have recently attracted interest, driven by the promise of new insights into synthetic topological phases of matter and Floquet states. This effort has culminated in the discovery of Cooper multiplets, in which the splitting of a Cooper pair is enabled via a series of Andreev reflections that entangle four (or more) electrons. In this text, we show conclusively that multiplet resonances can also emerge as a consequence of the three terminal circuit model. The supercurrent appears due to the correlated phase dynamics at values that correspond to the multiplet condition $nV_1 = -mV_2$ of applied bias. The emergence of multiplet resonances is seen in i) a nanofabricated three-terminal graphene Josephson junction, ii) an analog three terminal Josephson junction circuit, and iii) a circuit simulation. The mechanism which stabilizes the state of the system under those conditions is purely dynamical, and a close analog to Kapitza's inverted pendulum problem. We describe parameter considerations that best optimize the detection of the multiplet lines both for design of future devices. Further, these supercurrents have a classically robust $\cos2\phi$ energy contribution, which can be used to engineer qubits based on higher harmonics.
\end{abstract}

\maketitle

\section{Introduction}\label{sec1}

\begin{figure*}
    \centering
    \includegraphics[width=2\columnwidth]{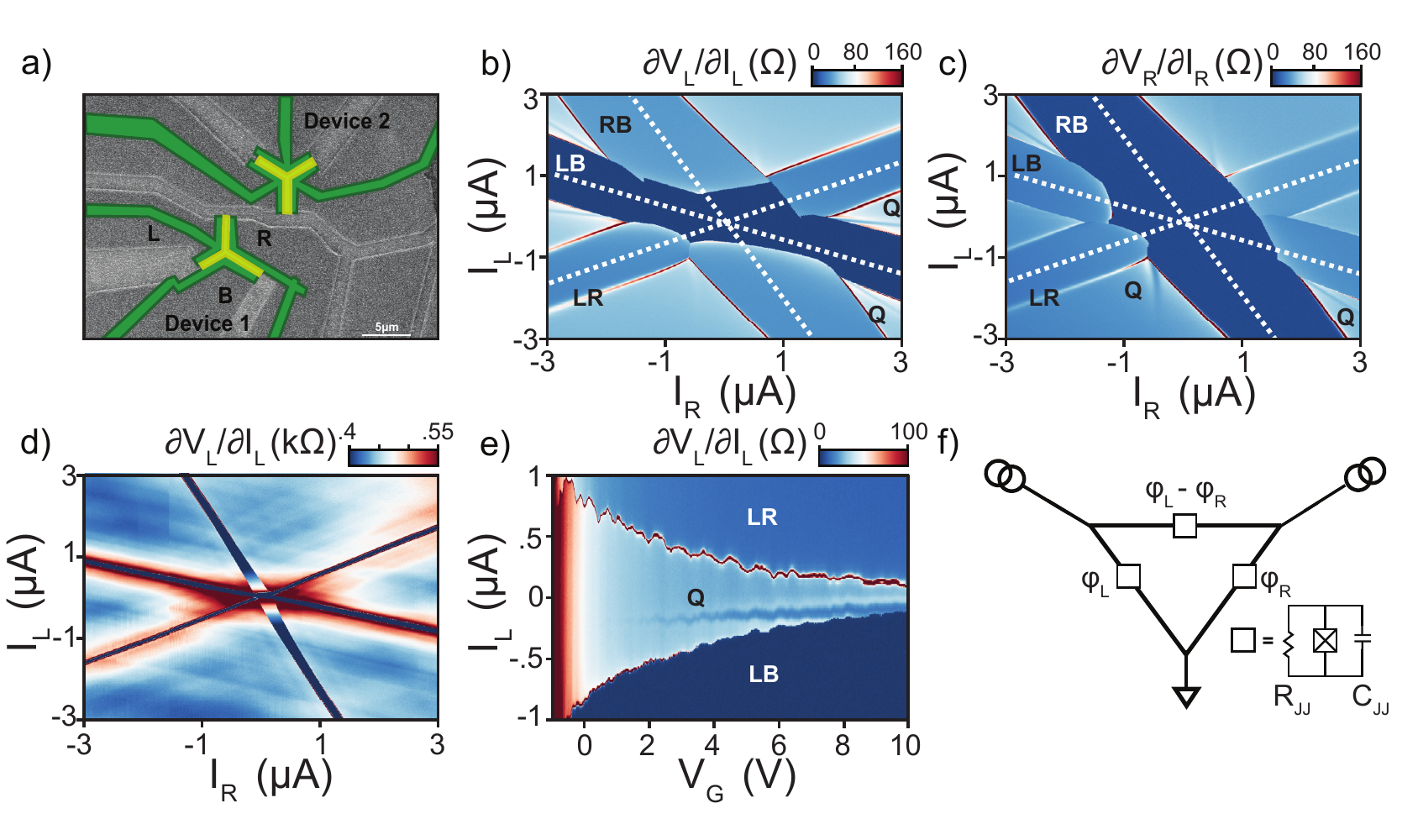}
 \caption {(a) SEM image of two three-terminal Josephson junctions patterned on an encapsulated BN/graphene device. Scale bar = 5 $\mu$m. (b) Differential resistance $\partial V_{L}/\partial I_{L}$ and c) $\partial V_{R}/\partial I_{R}$ as a function of both biases at $V_{G}=10$ V. Three large superconducting branches emerge corresponding to supercurrent between each pair of contacts. Additionally, a quartet branch (labelled Q) appears for each combination of $V_i=-V_j$. d) $\partial V_{L}/\partial I_{L}$ at $V_{G}=-2$ V. Near the Dirac peak, the Q lines disappear and dissipative MAR lines parallel to the superconducting branches emerge. e) $\partial V_{L}/\partial I_{L}$ for a fixed $I_R=2$ $\mu$A as a function of gate and $I_L$. The Q reduces in visibility as the resistance is increased. Variations in the position of the branch are due to electron interference in the ballistic cavity. f) The circuit model of the device. All three superconducting contacts are connected via  Josephson  junctions,  resistors  and  capacitors.}
    \label{fig:FigGraphene}
\end{figure*} 

A two terminal Josephson junction can be well described by the RCSJ model, whereby an imaginary particle representing the phase variable evolves in a tilted washboard potential~\cite{Tinkham}. When the particle rests in a minimum of the washboard, the phase is static and the device is superconducting. As a bias is applied, the washboard  tilts, until eventually the phase particle enters the running state, resulting in a voltage between the contacts proportional to $\langle\dot{\phi}\rangle$.

When a third superconducting contact is added, the washboard potential becomes two-dimensional, which expands the complexity of the phase trajectories~\cite{Arnault2021}. When a bias is applied to one of the contacts, it causes movement along the associated phase axis. If the phase is evolving along one axis but not the other, the junction associated with the stationary phase will generate a supercurrent, while the other junctions will develop a finite voltage \cite{Draelos2019,Pankratova2020,Pribiag2020}. 

Stationary phase conditions may arise even when the voltage across every pair of contacts is finite. For example, in a three terminal device such as the one shown in Figure 1a, when $V_{L}=-V_{R} \neq 0$,  $\langle \phi_{L}+\phi_R\rangle$ is stationary with respect to the grounded bottom contact. The microscopic origin of the resulting supercurrent has commonly been attributed to ``quartets" -- an entangled set of four electrons \cite{Pfeffer2014,Cohen2018,KoFan2021,graziano2022}. Without loss of generality, supercurrents generated by the static phase states exist for any combination of $nV_L+mV_R=0$, with integers $n$ and $m$ and involve the entanglement of multiplets consisting of four or more electrons. 

These multiplets have been an area of intense theoretical \cite{Nowak2019,MelinFloq,MelinBerry, SplitQuartet,QuartPiShift, Melo2021} and experimental \cite{Cohen2018,Pfeffer2014,KoFan2021} study as they require the multi-contact phase coherent trajectories necessary to realize synthetic topological states predicted in these devices \cite{Riwar2016, Eriksson2017, Meyer2017,Xie2017,Xie2018,Gavensky2018,GeomtricTensor, Strambini2016,Vischi2017}. Further, the multiplets themselves show promise to provide insights into Floquet dynamical systems \cite{KoFan2021,MelinFloq,MelinBerry, SplitQuartet,QuartPiShift}. However, it has recently been found that quartets may classically emerge as a consequence of the mixing of strictly sinusoidal current phase relations (CPR) \cite{Melo2021}.
Here, we demonstrate classical quartets in the case of a three terminal graphene Josephson junction (Figure 1a), a three-terminal Josephson junction analogue (Figures 2,3), and simulations of the multi-terminal circuit (Figures 4,5). 

\section{Results}\label{sec2}

\subsection{Graphene Three Terminal Junction}

The primary device studied here is Y-shaped (Figure 1a), with a 0.5 $\mu$m long graphene channel separating the three superconducting contacts of widths between 6.5 and 7.5 $\mu$m. The contacts are labelled left ($L$), bottom ($B$), and right ($R$), forming three junctions. The device length is comparable to the coherence length induced in graphene by MoRe \cite{Calado2015,Borzenets2016}, placing it in the intermediate length regime, where harmonics in the CPR should be relatively small \cite{Nadya2015PRB,Nanda2017}. This design should minimize any crossed Andreev reflections or multiple Andreev reflections between more than two contacts, as these processes could only occur in the relatively small central region where the three junctions meet. Also, the carriers would need to traverse the sample multiple times in order to be reflected from several contacts. Additional fabrication details are provided in Ref. \cite{suppinfo}. 

The device is cooled to a base temperature of 30 mK. We apply current biases from the L and R contacts, while measuring differential resistances from both leads to ground (B). We find results consistent with previous studies \cite{Draelos2019,Pankratova2020,Pribiag2020} (Figure 1b,c). Large superconducting branches correspond to supercurrent between each pair of contacts. Additionally, we observe quartet resonances at all combinations of $V_i=-V_j$ (labelled Q). 

Next, we vary the gate voltage on the device, depleting the graphene. Near the Dirac peak (Figure 1d), the contact transparency is significantly reduced, causing a simultaneous reduction of critical current and increase of $R_n$. In this region, the quartet lines disappear, while dissipative multiple Andreev reflection (MAR) lines emerge parallel to the primary superconducting branches. This observation is a strictly microscopic effect, consistent with Ref. \cite{Nowak2019}, and will not be captured by any of our modelling later on in the text. Theory and earlier work~\cite{Pfeffer2014} predict the emergence of quartet resonances due to nondissipative MAR between multiple transparent contacts. We believe, for several reasons, this is an unlikely explanation of the quartet resonances in our device. First, the distance between the contacts is too large to support even two-terminal higher harmonic supercurrents, which require carriers traversing between the contacts multiple times \cite{Arnault2021}. Indeed, multiplet resonances are robust to elevated temperatures \cite{InAs} (see \cite{suppinfo}). Second, the multi-terminal MAR trajectories resulting in quartets would be confined to the central area of the sample ($< 1\mu$m) and should be even further suppressed. Instead, as we will discuss later, the resonances we observe are a dynamical effect that is expected from the RCSJ model. 
Specifically, the variation of $R_n$ and $I_c$ is directly responsible for the loss of quartet lines and not the microscopics pertaining to nondissipative nonlocal MAR.

In Figure 1e, we fix $I_R$ at 2 $\mu$A, while sweeping $I_L$ and varying the gate voltage. The critical currents of the $LR$ and $LB$ junctions oscillate as a function of the gate voltage, which we attribute to electron interference within the ballistic cavity. These variations in critical current shift the location in current bias of the $V_L=-V_R$ condition and are reflected in variations of the quartet line. For this value of $I_R$, the quartet line is no longer visible near $V_{G}=1$ V. 
\begin{figure*}[htp]
    \centering
    \includegraphics[width=2\columnwidth]{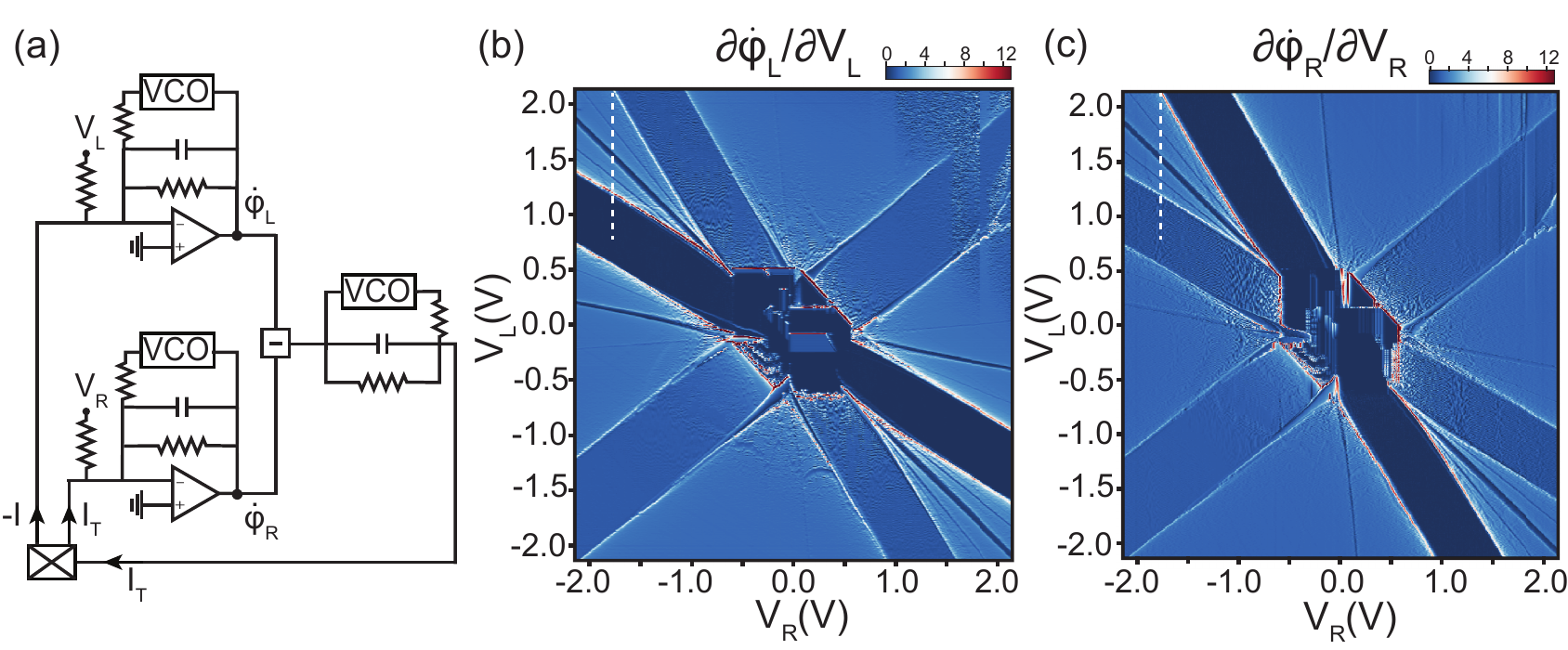}
 \caption {a) Simplified schematic of the analog three terminal junction. b) Map of the effective differential resistance $\partial \dot{\phi}_{L}/\partial V_{L}$ as a function of biases $V_{L}$ and $V_{R}$. c) Map of the effective differential resistance $\partial \dot{\phi}_{R}/\partial V_{R}$ as a function of both biases.  }
    \label{fig:FigA1}
\end{figure*} 

We model the three terminal junction by the current biased network of three shunted junctions shown in Figure 1f. The biased left and right contacts have superconducting phases $\phi_{L}$ and $\phi_{R}$, while the bottom contact is grounded with a phase of zero. Using Josephson equations and Kirchhoff laws, we can write the system of differential equations that the two phases obey. These can be cleanly written in matrix form  if one defines $\Phi=\begin{pmatrix} \phi_{L} \\ \phi_{R} \end{pmatrix}$ and $I=\begin{pmatrix} I_{L} \\ I_{R} \end{pmatrix}$. $\Phi$ can be shown to follow this differential equation:
\begin{equation}
\frac{\hbar}{2e}\mathcal{C} \ddot\Phi +\frac{\hbar}{2e}\mathcal{G}\dot\Phi + I_c(\Phi)=I
\end{equation}
$\mathcal{C}$ and $\mathcal{G}$ are 2x2 matrices that depend on the junctions shunting resistances and capacitances. $I_c(\Phi)$ depends on the junction's current phase relation, which for simplicity are assumed to be sinusoidal\cite{suppinfo}. In the rest of the paper, we develop two approaches to understand the dynamical properties of that equation. We first study a classical analog circuit that verifies the same equation, then turn to a powerful numerical scheme to solve it. 

\begin{figure*}[htp]
    \centering
    \includegraphics[width=2\columnwidth]{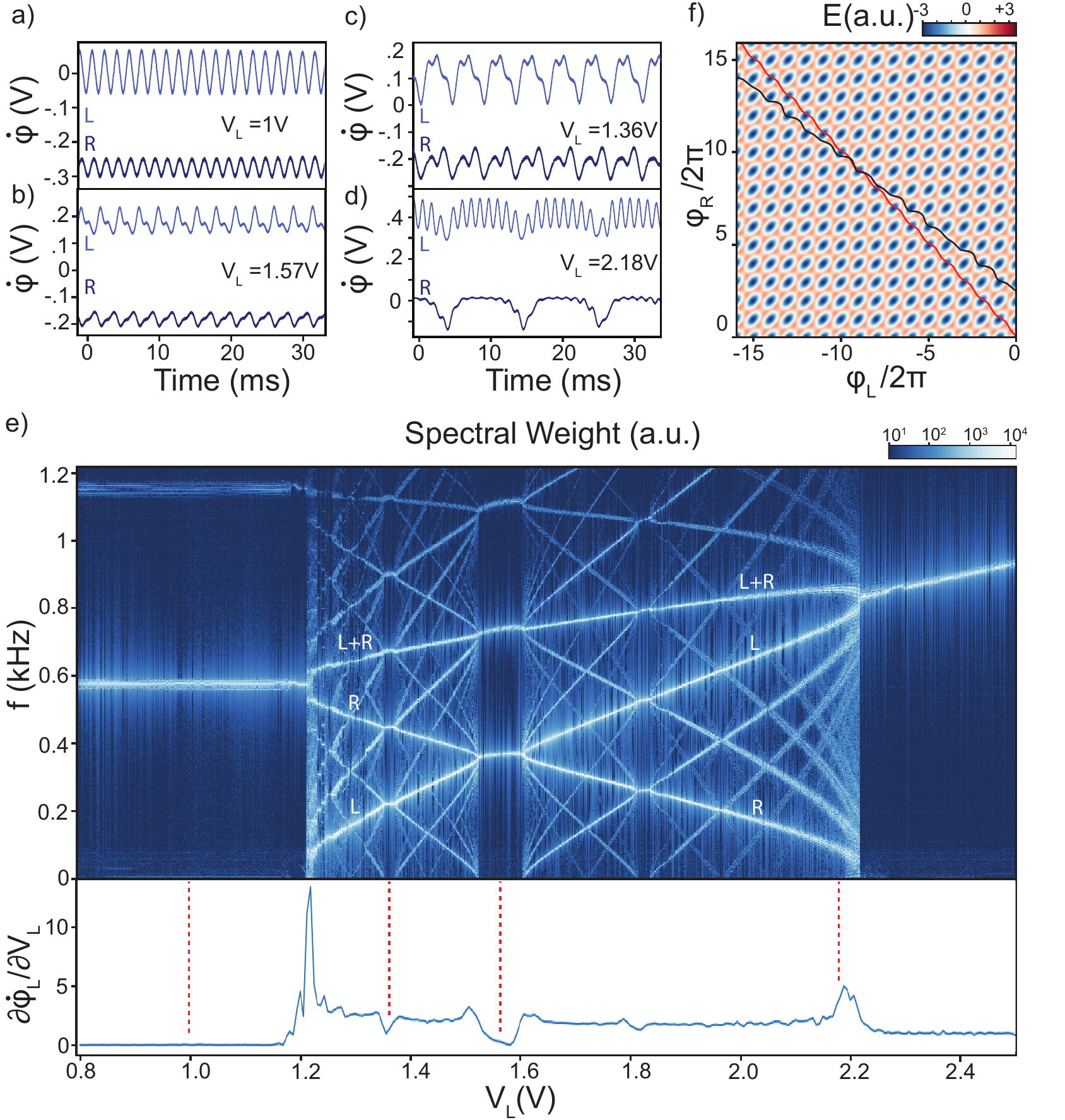}
 \caption {(a-d) Time evolution of $\dot{\phi}_{L}(t)$ and $\dot{\phi}_{R}(t)$, respectively in light and dark blue, measured at bias values along the vertical dashed line in  Figure 2c. These plots are obtained: (a) when junction LB is in the zero voltage state ($V_{L}$=1 V), (b) on top of the quartet resonance ($V_{L}$=1.57 V), (c) on top of one of the sextet resonances ($V_{L}$=1.36 V), and (d) close to the transition to the RB zero voltage state ($V_{L}$=2.18 V). (e) Map of the frequency spectrum of $\dot{\phi}_{L}(t)$ as a function of the bias $V_{L}$ from 0.8 V to 2.5 V, while the other bias $V_{R}$ is held constant at -1.8 V (white dashed line in Figure 2c). A time trace of $\dot{\phi}_{L}(t)$ is acquired with an oscilloscope at each bias value, and the fast Fourier transform is then calculated for that bias. The fundamental frequency of the left junction is constant at 600 Hz when it is in the zero voltage state up to $V_{L}\approx 1.2$ V, but it then increases monotonically with bias up to 1000 Hz as the junction enters the resistive state~\cite{Amet2021}.  Period doubling and tripling are evident on the quartet and sextet branch respectively. Bottom panel: Cross section of the differential resistance $\partial \langle\dot{\phi}_{L}\rangle/\partial V_{L}$ for the same bias values. f) Simulation of trajectories in phase space, superimposed on the washboard potential. Trajectories are shown on the quartet resonance (red) and away from it (black).}
    \label{fig:FigFFT}
\end{figure*} 

\subsection{Analog Three Terminal Junction}\label{sec2}

An analog circuit whose dynamical properties are identical to those of a three terminal junction is shown in Figure 2a \cite{dhumieres_chaotic_1982, blackburn_circuit_2007,hamilton_analog_1972,magerlein_accurate_1978}. The two main observables are the main operational amplifiers' outputs which are proportional to $\dot{\phi}_{L}$ and $\dot{\phi}_{R}$. Voltage-controlled oscillators provide the sinusoidal nonlinearity: if a voltage proportional to $\dot{\phi}$ is applied to their input, the output scales like $\sin(\phi)$. It is thus possible to show that $\phi_{L}$ and $\phi_{R}$ follow a system of differential equations formally identical to (1)\cite{Amet2021, suppinfo}. The circuit is therefore a classical implementation of the three terminal RCSJ model, and is expected to demonstrate the same dynamical effects as a multi-terminal junction, but free from any microscopic ``artifacts". We show in Ref. \cite{suppinfo} that the two input voltages $V_{L}$ and $V_{R}$ play the same role as input currents $I_{L}$ and $I_{R}$ in Figure 1f. 
 
In a conventional junction, the oscillations of $\dot{\phi}$ occur on sub-nanosecond time-scales. This dynamic is inevitably time-averaged in conventional transport measurements by the slower response of the setup, and in practice only the DC component $\langle \dot{\phi} \rangle$ is measured. This is not the case in our analog circuit, since phases evolve on ms time-scales and can therefore be recorded. This junction analogue thus provides a wealth of information on phase dynamics which is experimentally inaccessible in a conventional junction. 

We first discuss the time-average of $\dot{\phi}$ to compare it to previously described Josephson junction measurements. The amplifier outputs $\dot{\phi}_{L}$ and $\dot{\phi}_{R}$ are time-averaged with low-pass filters. Since $V_{L}$ and $V_{R}$ mathematically play the same role as current biases, we plot the quantities $\partial \dot{\phi}_{L}/\partial V_{L}$ and $\partial \dot{\phi}_{R}/\partial V_{R}$ which are formally equivalent to the differential resistances of the left and right junction, even though these quantities are technically dimensionless. We obtain the maps presented in Figure 2b and 2c, which are strongly reminiscent of prior transport measurements on three terminal junctions. 

Indeed, we notice the three widest diagonal arms of suppressed differential resistance, which each correspond to one of the analog junctions being locked into the zero-voltage state. In Figure 2b, the darkest branch corresponds to the LB junction, whereas in Figure 2c, it corresponds to the RB junction. Remarkably, the data also show narrower resonances of suppressed differential resistance when $\langle \dot{\phi}_{L}+\dot{\phi}_{R}\rangle=0$, $\langle \dot{\phi}_{L}+\dot{\phi}_{T}\rangle=0$, and $\langle \dot{\phi}_{R}+\dot{\phi}_{T}\rangle=0$. These correspond to quartet resonances which are obviously not a microscopic effect and solely stem from classical phase dynamics. Note that the circuit even generates classical sextet resonances along lines such as $\langle \dot{\phi}_{L}+2\dot{\phi}_{R}\rangle=0$ and $\langle 2\dot{\phi}_{L}+\dot{\phi}_{R}\rangle=0$. 

To further explore these classical multiplet resonances, 
we turn to the non-averaged time evolution of $\dot{\phi}_{L}$ and $\dot{\phi}_{R}$. For a three-terminal junction, the washboard potential is:
\begin{equation}
U(\phi_L,\phi_R) = - E_{L} \cos(\phi_L)- E_{R} \cos(\phi_R) - E_{LR} \cos(\phi_L-\phi_R)
\end{equation}
$E_{L}$, $E_{R}$ and $E_{LR}$ represent the Josephson energies of the left, right and transverse junctions. $U(\phi_L,\phi_R)$ is plotted on Figure 3f in the particular case where the three Josephson energies are identical, which turns out to be the case in the analog circuit presented here. If the left junction is in the zero-voltage state, its phase oscillates around zero and the trajectory would be nearly vertical on Figure 3f. Figure 3a shows the time evolution of the $\phi_{L}(t)$ and $\phi_{R}(t)$ in this scenario: $\dot{\phi_L}$ oscillates around 0, while $\langle \dot{\phi_R}\rangle \neq 0$. Similarly, if the right junction were in the zero voltage state, $\phi_{R}(t)$ would oscillate around zero and the trajectory would be nearly horizontal in Figure 3f. 

We now turn to the phase dynamics when none of the three junctions are in their respective zero-voltage state. Figure 3e represents the frequency spectrum of $\dot{\phi}_{L}(t)$ as a function of the bias $V_{L}$, but the corresponding map for $\dot{\phi}_{R}(t)$ is nearly identical. $V_{R}$ is held constant at -1.8 V. This corresponds to a vertical cut on Figure 2b where the device starts in the zero-voltage arm LB (at $V_{L}=0.8$ V), then cuts through dissipative regions and multiplet resonances before it ends in the zero-voltage arm RB (at $V_{L}=2.5$ V). In order to acquire the frequency spectrum, a 1.4 s time trace of $7\times 10^{5}$ points is acquired at every bias value with an oscilloscope. We then compute the fast Fourier transform at each bias value to obtain the map 3e.  

For $V_{L}\leq 1.2$ V, the left junction is in the zero voltage state while the right junction is in the running state. $\dot{\phi}_{R}(t)$ thus oscillates with a non-zero average, as shown in Figure 3a. Those oscillations are also seen in $\dot{\phi}_{L}(t)$ as a result of the coupling between the two junctions. In that regime, the oscillation frequency only depends on the bias in the $R$ direction, and the resonance seen at $580$ Hz on Figure 3e is therefore flat until $V_{L}=1.2$ V.  When $V_{L}$ exceeds 1.2 V, $\phi_{L}$ enters the running state and new resonances appear. The drift of the phase in the $L$ direction is very slow at first, but speeds up with increasing $V_{L}$. It therefore results in a frequency component in the spectrum, labeled $L$, which starts close to 0 Hz at $V_{L}=1.2$ V and increases up to $\approx 800$ Hz at $V_{L}=2.2$ V [Figure 3e]. Meanwhile, as $V_{L}$ approaches 2.2 V the right junction gets closer to its zero-voltage state and oscillations caused by $\dot{\phi}_{R}$ slow down. This corresponds to the frequency component that decreases from 580 Hz at 1.2 V to 0 at 2.2 V, labeled $R$. Very close to 2.2 V, the trajectory along the washboard potential in Figure 3f is nearly horizontal, but phase-slips in $\phi_{R}$ occur every few oscillations of $\phi_{L}$, which causes a spike in $\dot{\phi}_{R}$. The oscillations of $\dot{\phi}_{L}$ are thus expected to be modulated at a very slow frequency by the ratcheting of $\phi_{R}$. This scenario is shown in Figure 3d, and it corresponds to a region of the FFT map close to the right edge of the diffusive region ($\approx 2.2$ V).
Finally, we note that the nonlinear coupling between the two junctions causes a third strong resonance at the sum of the first two frequencies, and labeled $L+R$. Note that other frequency combinations, such as $L-R$ and $R-L$ are also generated but not labeled, so as not to crowd the map.

Multiplet resonances emerge when the $L$ and $R$ frequency components are commensurate. For example, when $V_{L}\approx 1.55$ V, $L=R$ and the cut intersects the classical quartet resonance in 2b. An excerpt of the corresponding time trace is shown in Figure 3b. Here $L+R$ simply corresponds to a frequency doubling of the main resonance, which explains the double-peaked profile of the time trace. Similarly, sextet states are observed when $L=2R$ or $R=2L$, around biases 1.35 V and 1.85 V. A relevant time-trace is shown on Figure 3c. 
For both quartet and sextet resonances, the commensurate condition persists over a finite range of bias $V_{L}$, which explains the finite width of those resonances in 2b. We prove this result analytically in Ref.\cite{suppinfo} in the quartet case. 
\section{Simulations} \label{subsec2}

We now turn to numerical simulations of Equation (1). We chose to study the multi-terminal generalization of the conventional RCSJ model, rather than to use the full model which takes into account the lead resistance and the capacitance of the bonding pads~\cite{Trevyn}. Indeed, the full model effectively reduces to RCSJ model in our range of parameters (high critical currents). Furthermore, the analog system studied in Figures 2 and 3 directly corresponds to RCSJ without extra circuit elements.

Consistent with previous experimental work, we find superconducting branches corresponding to supercurrents between each pair of contacts for $V_i=0$ (Figure 4a,b). We also observe the additional multiplet resonant branches at voltage values corresponding to V$_i=-V_j$ and V$_i=-2V_j$. These resonances are thus confirmed to be a purely dynamical effect that results from the RCSJ model. 

To understand the dynamics, we plot the trajectory in phase space when biases are such that the device is in the quartet dynamical state I (in red on Figure 3f). Understandably, the overall trajectory follows a contour of constant $\phi_{L}+\phi_{R}$, although oscillations along that contour are noticeable. Note that this trajectory is calculated slightly off the the center of the quartet resonance to better emphasize its stability. When the bias drives the junction out of the classical quartet resonance, the trajectory in phase space loses its symmetry and consists of more random phase jumps in the $\phi_{L}$ and $\phi_{R}$ direction (in black on Figure 3f). Intuitively, one expects the red trajectory to be robust against small perturbations in the bias. 

\begin{figure}
    \centering
    \includegraphics[width=\columnwidth]{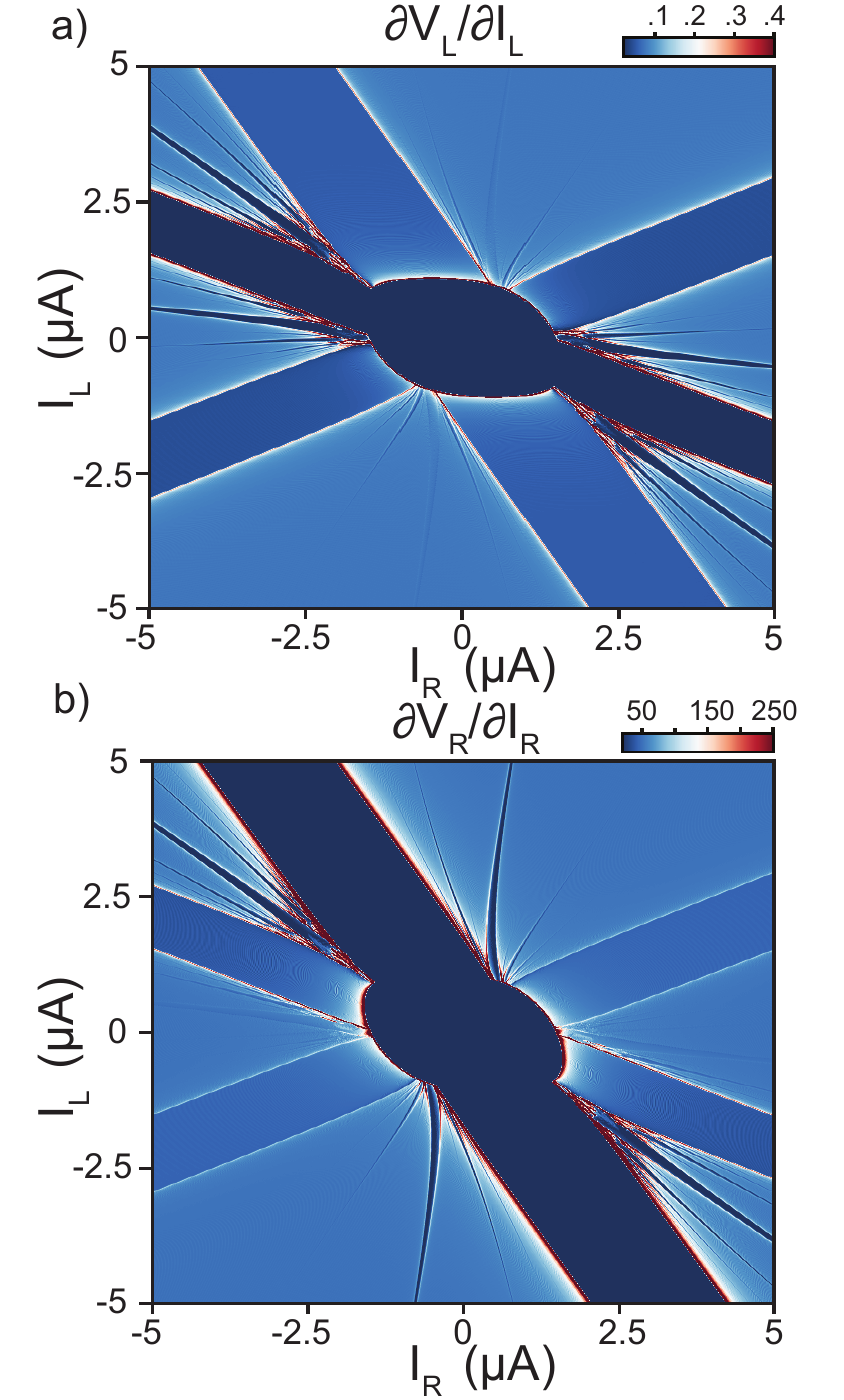}
    \caption {Numerical simulation of the differential resistance $\partial V_{L}/\partial I_{L}$ (panel a) and $\partial V_{R}/\partial I_{R}$ (panel b) as a function of both biases.   Primary superconducting branches emerge corresponding to supercurrents between each pair of contacts. Additional resonances emerge showing quartet and higher multiplet supercurrents.}
    \label{fig:Figsim}
\end{figure}
Along the quartet resonance, the phase space trajectory approximates $\phi_L +\phi_R \equiv 0\mod2\pi$. The potential energy along that cross section can thus be rewritten as:
\begin{align*}
U(\phi) \approx - (E_{L}+E_{R}) \cos(\phi_{L,R})
- E_{LR} \cos(2\phi_{L,R}) 
 \hspace{1mm}
\end{align*}

This potential energy has two local maximas per period. It is therefore understandable that the angular velocity $\dot{\phi}$ has a double-peak profile for both channels in Figure 3b.


While multiple higher order resonances are observed in the simulation, experimentally these resonances are extremely weak (Figure 2). We attribute this suppression to the sub-optimal tuning of the circuit elements. Indeed, the stability of these multiplet resonances strongly depends on circuit parameters, and in particular on the quality factor of the junctions. 

\section{Discussion}

The stability of the quartet resonance and its dependence on circuit parameters can be understood analytically within the framework of equation (1). If, for simplicity, all shunting capacitances and resistances are assumed to be identical, we can define $\epsilon=\frac{\phi_{L}+\phi_{R}}{2}$,  $\eta=\frac{\phi_{L}-\phi_{R}}{2}$, $I_{+}= I_{L}+I_{R}$, $I_{-}= I_{L}- I_{R}$, and show that:

\begin{equation}
    \ddot{\epsilon}+\frac{\omega_{0}}{Q}\dot{\epsilon}+\omega_{0}^{2}\cos(\eta)\sin(\epsilon)=I_{+}
\end{equation}

\begin{figure*}
    \centering
    \includegraphics[width= 2\columnwidth]{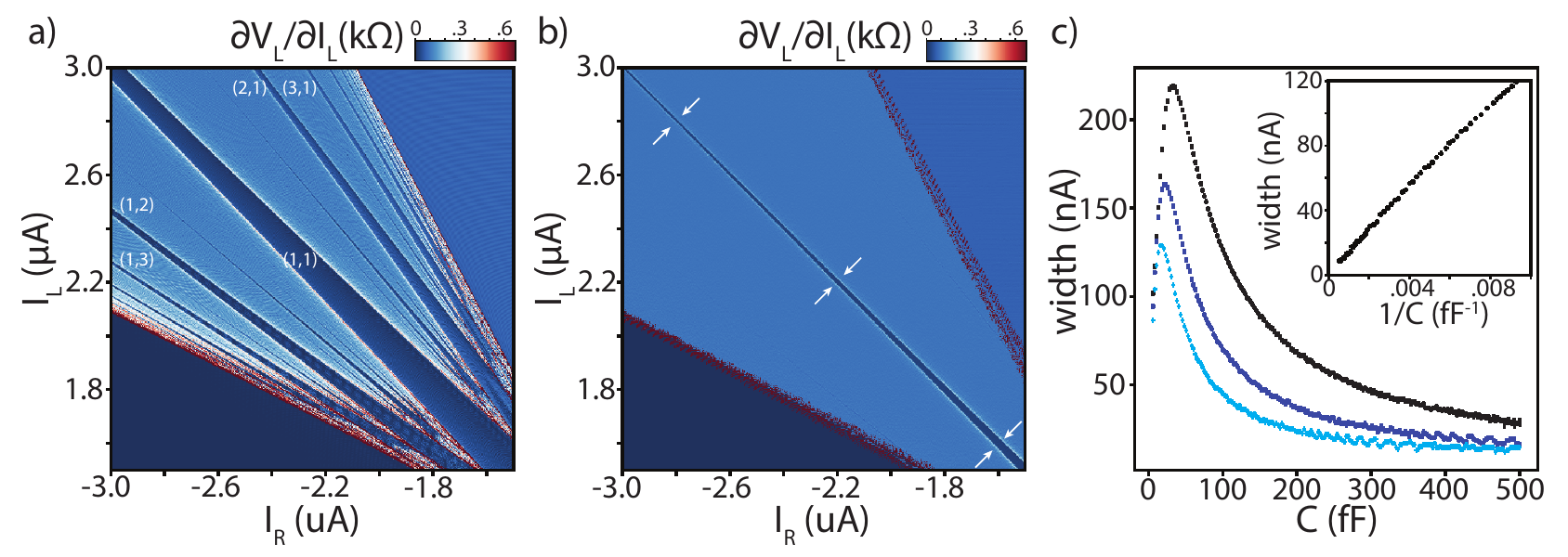}
    \caption {a,b) Numerical simulation of the differential resistance $\partial V_{L}/\partial I_{L}$ vs. biases $I_{L,R}$, which are limited to the top left quadrant. For simplicity, all parameters are kept identical for the three junctions: critical currents are $Ic = 600$ nA, shunting resistances are $R =160 \Omega$, shunting capacitances $C$ are all equal to a) 30 fF in panel (a), and 300 fF in panel (b). c) The width of the quartet resonance $\delta I$ as a function of $C$. The widths are measured at biases indicated by white arrows in panel b: $I_{L}-I_{R}=3.2 \mu$A (black), $4.4\mu$A (dark blue), and $5.6 \mu$A (light blue). Inset: Width of the resonance for $I_{L}-I_{R}=3.2\mu$A as a function of $1/C$.}
    \label{fig:Figs0}
\end{figure*}

Close to the quartet resonance, $\eta\approx\phi_{L,R}$ changes very rapidly and is in fact nearly linear in time: $\eta \approx \omega t$ with $\omega\equiv\frac{e(V_{L}-V_{R})}{\hbar}$. This approximation makes equation (3) equivalent to Kapitza's inverted rigid pendulum problem (in the absence of gravity). The rapid oscillations of $\cos(\eta)$ stabilize $\epsilon$ near an equilibrium at $\epsilon =0$ or $\pi$, and therefore lock the phase around a contour of constant $\phi_{L}+\phi_{R}\equiv 0\mod2\pi$, which is the quartet trajectory in phase space. With minimal changes to the canonical solution of Kapitza's problem, we show that $\langle \dot \epsilon \rangle$ stays zero  for a small, but nonzero range of $ \lvert I_{+} \rvert < \delta I$, with

\begin{equation}
   \delta I = \frac{\hbar I_{C}^{2}}{4eC(V_{L}-V_{R})^{2}}
\end{equation}

Within this range, the corresponding differential resistance, $\propto d \langle\dot{\epsilon}\rangle/d I_{+}$, is equal to zero, which explains why those classical multiplet trajectories result in superconducting branches in the differential resistance maps \cite{suppinfo}.


In Figures 5a and 5b we plot the simulated differential resistance $\partial \dot{\phi}_{L}/\partial I_{L}$ as a function of $I_{L,R}$ for two values of capacitance, $C=30fF$ and $C=300fF$. Multiplet resonances are very clearly seen in 5a when $p\dot{\phi}_{L}+q\dot{\phi}_{R}=0$, with $(p, q) \leq 5$. We labeled the first few values of $(p,q)$ in white. However, those resonances are heavily suppressed at higher capacitances, Figure 5b.

We next simulate the width of the quartet resonance $\delta I$ for three different values of bias ($I_-= I_L - I_R$) over a wide range of $C$. As expected, $\delta I$ decreases with bias. The width is also found to be proportional to $1/C$ at high $C$ (Figure 5c), in agreement with eq.~(4). Interestingly, $\delta I$ is nonmonotonic in $C$ and reaches a local maximum at very low capacitance. This trend is not captured by the derivation above, which required a clear separation of the time scales $\omega>>\omega_{0}$. When the capacitance drops, $\omega_{0}$ becomes too large for this condition to hold. Indeed the position of the local maximum shifts to lower capacitance values when $\omega$ increases with bias.

Our results show that oscillator synchronization can produce stable supercurrents while all junctions in the device are in the running state. This counterintuitive result provides evidence that multi-terminal Josephson junctions may host a number of macroscopic quantum phenomena (such as a supercurrent) based solely on the classical nonlinear equations that dictate their dynamics. This concept may prove to be useful as Josephson junctions rise to prominence as a fundamental building block of quantum computers.

More specifically, the robust classical 2$\phi$ periodicity of the multiplet resonances may be useful in developing $\cos 2\phi$ qubits \cite{Gladchenko2009,Smith2020,Melo2021}. In this case, coherence could be generated using flux biasing with superconducting loops, which have been used to probe topological states in diffusive multiterminal junctions \cite{Strambini2016,Vischi2017}. The robust $\cos 2\phi$ energy can then be achieved by appropriately varying the contact phase through the $\phi_L=-\phi_R$ condition.

\section{Methods}

Graphene and boron nitride flakes are separately exfoliated on a thermally oxidized silicon substrate. The graphene is then encapsulated between BN layers using a dry-transfer method, and deposited on a doped silicon substrate with a 280 nm thick oxide. This protects the sample against contaminants and allows for ballistic transport \cite{Dean2010, Mayorov2011}, including ballistic supercurrent over several microns \cite{Calado2015,BenShalom2016,Borzenets2016}.
The structure is then annealed in atmosphere at 500ºC for one hour. The device region is defined using electron beam lithography and is etched using a CHF$_3$ / O$_2$ reactive ion etching process. The three superconducting electrodes consist of 70 nm thick molybdenum rhenium alloy, a material known to make high transparency contacts to graphene \cite{BenShalom2016,Borzenets2016}. The MoRe is sputtered at 70W in an argon atmosphere at a pressure of 3 mTorr, and directly after a reactive ion etch.

The device is cooled in a Leiden Cryogenics dilution refrigerator and measured using standard lock-in techniques.

Numerical simulations involve a fourth-order Runge-Kutta scheme written in tensor form so that the computation of $\Phi(t)$ at all bias points can be parallelized over a large number of GPU cores in PyTorch \cite{Arnault2021,suppinfo}. Maps shown in Figure 4 are thus generated in under 20 s \cite{suppinfo}. Once the time evolution $\Phi(t)$ is determined, we compute the time-average of $\dot{\Phi}(t)$ to determine DC voltages across each junctions. This allows us to compute the differential resistances $\partial \dot{\phi}_{L}/\partial I_{L}$ and $\partial \dot{\phi}_{R}/\partial I_{R}$ as a function of both biases (Figure 4a,b). Additional details on the numerical scheme are shown in Ref.\cite{suppinfo}.



\section*{Acknowledgments}
F.A. thanks Jacob Gardner for introducing him to parallel computing in Pytorch. We also thank Brian Opatosky, Wade Hernandez, and Patrick Richardson for their technical input. We thank Anton Akhmerov and Valla Fatemi for useful conversations. Transport measurements of graphene samples by E.G.A. and T.F.Q.L., and data analysis by E.G.A., L.Z. and G.F., were supported by Division of Materials Sciences and Engineering, Office of Basic Energy Sciences, U.S. Department of Energy, under Award No. DE-SC0002765. Lithographic fabrication and characterization of the samples performed by E.G.A., F.A., and L.Z. were supported by the NSF Award DMR-2004870. S.I. was supported by a GRAM fellowship. F.A., and A. M. were supported by a URC grant at Appalachian State University.  K.W. and T.T. acknowledge support from JSPS KAKENHI Grant Number JP15K21722 and the Elemental Strategy Initiative conducted by the MEXT, Japan. T.T. acknowledges support from JSPS Grant-in-Aid for Scientific Research A (No. 26248061) and JSPS Innovative Areas “Nano Informatics” (No. 25106006). This work was performed in part at the Duke University Shared Materials Instrumentation Facility (SMIF), a member of the North Carolina Research Triangle Nanotechnology Network (RTNN), which is supported by the National Science Foundation (Grant ECCS-1542015) as part of the National Nanotechnology Coordinated Infrastructure (NNCI). 

\newpage

\section*{Supplementary materials}

\subsection{Heating}

We showed in Figure 1 of the main paper that quartet resonances appear in the differential resistance measured on graphene-based three terminal junctions. Here we discuss the temperature dependence of those resonances. Indeed, temperature-dependent measurements are a means to distinguish microscopic mechanisms and macroscopic circuit effects that could generate quartet resonances. As the temperature is elevated, the Andreev bound state structure from higher harmonics in the CPR or nondissipative multiple Andreev reflection states should rapidly fade away \cite{InAs}. However, features that arise due to the larger energy scales of the circuit should be more robust \cite{Arnault2021}. 

We apply current biases from the left and right contacts while measuring the differential resistance of the left contact to ground. The gate was set to 0 V (approximately 2 V away from the Dirac peak.) We track the quartet resonance between the LB and LR branches and compare the visibility of the quartet resonance as we increase temperature. At base temperature (Figure S1a), the quartet line is clearly visible. As temperature is raised, there is a distinct reduction of visibility in the quartet line (Figure S1b,c). This culminates in the complete loss of quartet visibility at 1.75 K (Figure S1d). We are thus able to observe the quartet resonance at relatively elevated temperatures. This suggests that they result from the dynamical properties of the RCSJ model, and not from higher harmonics in the CPR.

\begin{figure}
    \centering
    \includegraphics[width= \columnwidth]{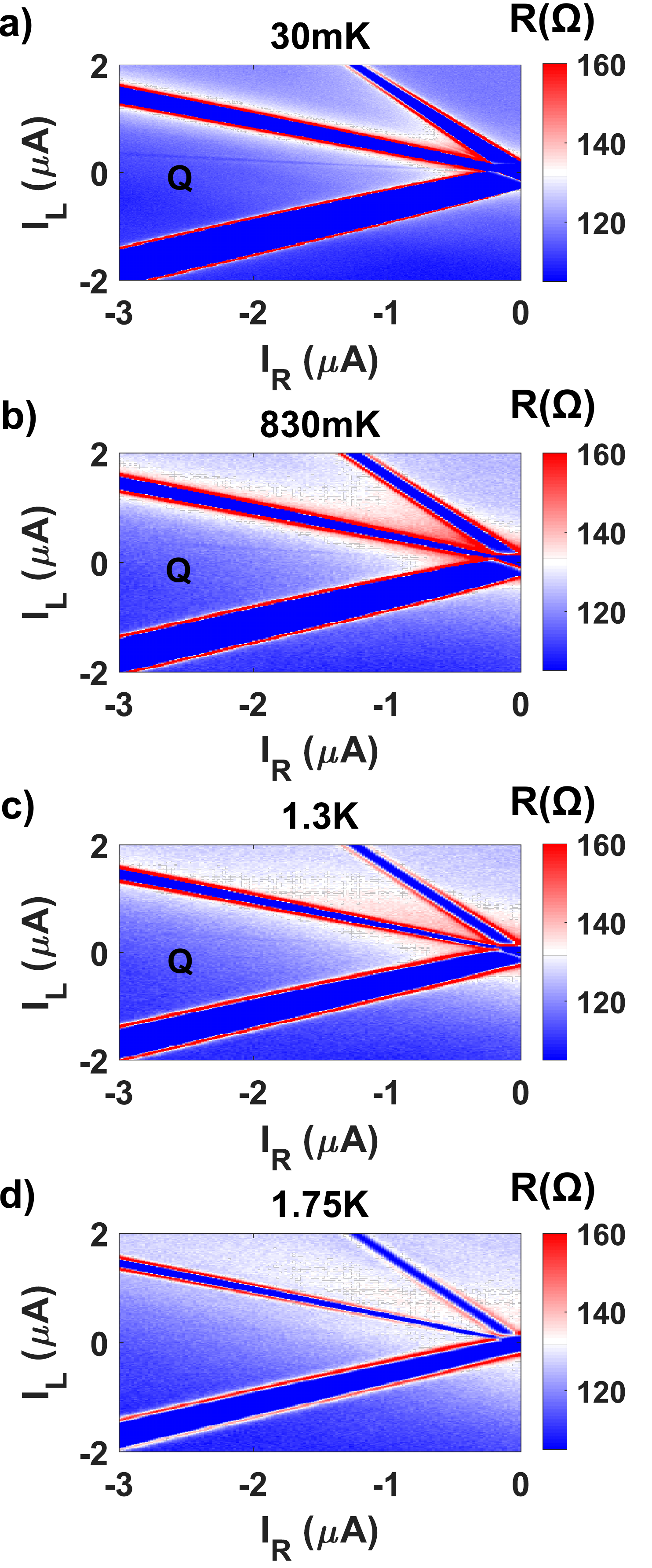}
 \caption {The effects of heating on the visibility of the quartet resonance. a) At the base temperature the resonance is clearly seen. b,c) As temperature is raised the resonance becomes more faint before d) the resonance disappears. The
 robustness to elevated temperature indicates that the origin of the resonance is not due to a complicated microscopic mechanism.}
\end{figure} 

\subsection{Shapiro Steps}

To show the coherence of the quartet resonances we apply 5 dBm of 5.2 GHz microwave radiation to our device. We can see the emergence of Shapiro steps, which result from the locking of the two superconducting phases onto the microwave drive. It generates quantized voltage steps at $V_n=\frac{nhf}{2e}$, where $n$ is typically an integer. Higher harmonics in the CPR generate additional steps at voltages $V_{n,m}=\frac{nhf}{2me}$, with both $n$ and $m$ as integers. $m$ represents the harmonic of the current phase relationship - $I(\phi)=\sum_m I_{c,m}\sin m \phi$. One would expect that, along the quartet resonance, the robust $\sin2\phi$ supercurrent contribution from the energy landscape would give rise to half integer steps. Alternatively, in the case of entangled Cooper quartets (corresponding to transport of a 4e charge) and barring any harmonics in the supercurrent contribution, the plateau values would take $V_n=\frac{nhf}{4e}$, which would also appear as half integer steps.

Unfortunately, the small supercurrents make quantitative discussion of the plateau values meaningless - the branches are too small to resolve the voltage step on top of the dissipative background. However, the existence of the Shapiro branches demonstrates that the multi-phase potential is coherent. We note however that this only means that the superconducting phases of the contacts are synchronized and implies nothing about the entanglement of the transport.

\begin{figure}
    \centering
    \includegraphics[width= \columnwidth]{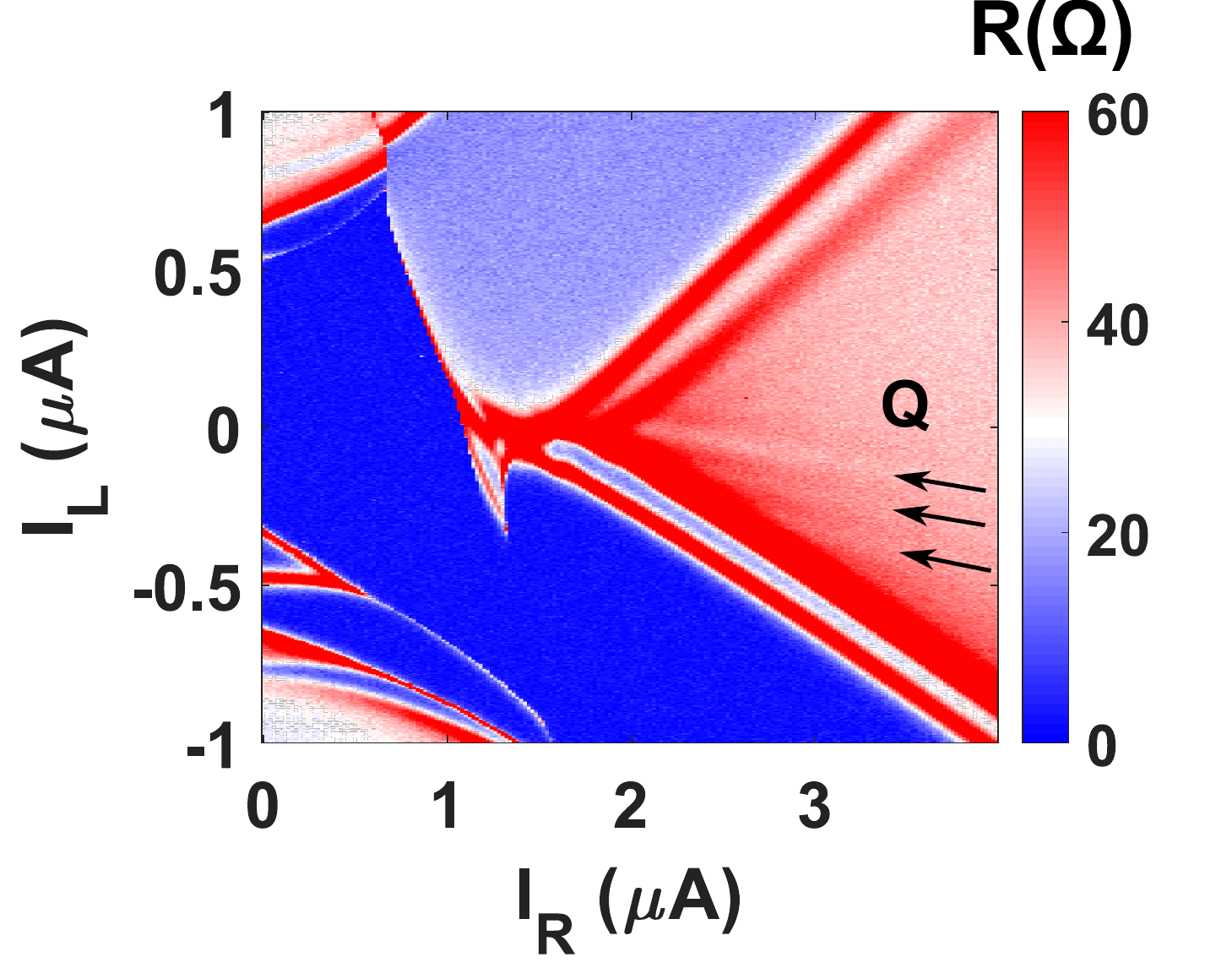}
 \caption {Bias-bias differential resistance map of the LB junction with 5 dBm of 5.2 GHz applied microwave  signal. Arrows marked  with ``Q” indicate the  Shapiro steps of the Quartet.}
    \label{fig:FigS1}
\end{figure} 

\subsection{Numerical scheme}

Equation (1) in the main paper was solved numerically using a fourth order Runge-Kutta scheme. The phase $\Phi(t)$ can be calculated for each value of bias sequentially, but that procedure is extremely slow and maps such as the ones shown in the main paper would take several hours to compute. Alternatively, we describe here how to rewrite the differential equation using tensor notation from the Python package Pytorch, so that the computation for all bias values can be done in parallel, and distributed over the GPUs of a graphic card. In our case, this procedure sped up the computation by over two orders of magnitude.  

First, we can rewrite equation 1 as a set of two first order differential equations:

\begin{align*}
\dot{\Phi}&=\Gamma\\
\dot{\Gamma}&=-\mathcal{C}^{-1}\mathcal{G}\Gamma+\frac{2e}{\hbar}\mathcal{C}^{-1}(I-I_{c}(\Phi))
\end{align*}

We have:
\begin{equation}
I_{c}(\Phi)=\begin{pmatrix}I_{L}\sin(\phi_{L})+I_{T}\sin(\phi_{L}-\phi_{R})\\I_{R}\sin(\phi_{R})-I_{T}\sin(\phi_{L}-\phi_{R})\end{pmatrix}
\end{equation}

We first note that since $\Phi$ and $\Gamma$ are both two row vectors, in the absence of the nonlinear term $I_{c}(\Phi)$, this system of differential equation could be rewritten as a single first order differential equation involving a $4\times 4$ matrix operating on a 4-row vector. 

We then define a third-order tensor $\Psi$(t) of dimension $[N_{i},N_{j},4]$ such that $\Psi[i,j,:]$(t) corresponds to the following four row vector for two specific values of the biases $I_{L}[i]$ and $I_{R}[j]$:
\begin{equation}
\Psi[i,j,:](t)=\begin{pmatrix}\phi_{L}(t)\\\phi_{R}(t)\\\dot{\phi}_{L}(t)\\\dot{\phi}_{R}(t)\end{pmatrix}
\end{equation}

The system of differential equations at all bias values can then be written as a set of tensor operations acting on $\Psi$ as a whole:

\begin{align}
\dot{\Psi}=\mathcal{F}(\Psi)
\end{align}

The function $\mathcal{F}$ operates on tensors of dimension $[N_{i},N_{j},4]$. For example, in the absence of the nonlinear term $I_{c}(\Phi)$, we would just write $\mathcal{F}(\Psi)=A\Psi+B$ where $\mathcal{A}$ is a tensor of dimension $[1,1,4,4]$ such that $\mathcal{A}[0,0]$ is a $4\times 4$ matrix operating on the 4-row vector $\Psi[i,j,:]$(t), and $\mathcal{B}$ is a $[N_{i},N_{j},4]$ constant tensor.

In the present case, however, the function $\mathcal{F}$ is nonlinear but it can still be written using tensor-compatible operations in Pytorch. For example, one can write:

\begin{equation}
\mathcal{F}(\Psi)=\mathcal{A} 
\Psi
-\frac{2e}{\hbar}\mathcal{\tilde{C}}^{-1}(\mathcal{B}\sin(\Psi)+I_{T}\sin(\mathcal{D}\Psi)-\mathcal{I})
\end{equation}

Where we used these notations:
\begin{align*}
    \mathcal{A}&=\left(
\begin{array}{c}
\begin{array}{c|c} 
  \begin{array}{c c} 
     0 & 0 \\ 
     0 & 0 
  \end{array}&
  \begin{array}{c c} 
  \hspace{4.5mm}
     1 & 0 \\ 
    \hspace{4.5mm}  0 & 1 
  \end{array}
\end{array}\\\hline
\begin{array}{c|c} 
  \begin{array}{c c} 
     0 & 0 \\ 
     0 & 0 
  \end{array}&
  -\mathcal{C}^{-1}\mathcal{G}
\end{array}
\end{array}
\right)\hspace{5mm}
\mathcal{B}=\begin{pmatrix}0 & 0 & 0 &0\\
0 & 0 & 0 &0\\
I_{L} & 0 & 0 &0\\
0 & I_{R} & 0 &0\end{pmatrix}\hspace{5mm}
\\
\mathcal{
\tilde{C}}^{-1}&=\left(
\begin{array}{c}
\begin{array}{c|c} 
  \begin{array}{c c} 
     0 & 0 \\ 
     0 & 0 
  \end{array}&
  \begin{array}{c c} 
     0 & 0 \\ 
     0 & 0 
  \end{array}
\end{array}\\\hline
\begin{array}{c|c} 
  \begin{array}{c c} 
     0 & 0 \\ 
     0 & 0 
  \end{array}&
  \mathcal{C}^{-1}
\end{array}
\end{array}
\right)\hspace{9mm}
\mathcal{D}=\begin{pmatrix}1 & -1 & 0 &0\\
-1 & 1 & 0 &0\\
0 & 0 & 0 &0\\
0 & 0 & 0 &0\end{pmatrix}
\end{align*}

\begin{align*}
    \mathcal{I}[i,j,:]=\begin{pmatrix}0\\0\\I_{L}[i]\\I_{R}[j] \end{pmatrix}
\end{align*}

Note that two singleton dimensions must be appended to the left of all $4\times 4$ matrices, to turn them into tensors of dimension $[1\times 1\times 4\times 4]$ which can operate on $\Psi$. Note that $\mathcal{I}$ is also a third order tensor of dimension [Ni,Nj,4]. In that equation, the "$\sin(\Psi)$" should be understood as the operation torch.sin in Pytorch which returns a tensor of identical dimension with the sine of each element. 

With those notations, the tensor $\Psi$ can thus be updated as a whole at each time step, which avoids two for-loops iterating over current bias values $I_{L}[i]$ and $I_{R}[j]$. The rest of the code is more akin to a conventional fourth-order Runge-Kutta scheme, but the tensor notation allows it to be parallelized over the GPU cores of the computer's graphic card ($\approx$ 4000 cores in our case). This speeds up the computation by over two orders of magnitude, and simulated maps shown in this work only take $\approx$ 20s to compute for a 400 by 400 pixel map.

The parameters that are used for the simulations presented in Figure 4 of the main paper are the following:
\\
\begin{center}
\begin{tabular}{ |c|c|c|c|c|c|c|c|c| } 
 \hline
 $I_{L}$ & $I_{R}$ & $I_{T}$ & $R_{L}$ & $R_{R}$ & $R_{T}$ & $C_{L}$ & $C_{R}$ & $C_{T}$ \\\hline  
 600 nA & 950 nA & 750 nA & 72 $\Omega$ & 32 $\Omega$ & 43 $\Omega$ & 50 fF & 50 fF & 50 fF \\ 
 \hline
\end{tabular}
\end{center}

\vspace{5mm}
In Figure 5, we chose symmetric circuit parameters in order to focus on the capacitance dependence of the width of the quartet resonance:
\\
\begin{center}
\begin{tabular}{ |c|c|c| } 
 \hline
 $I_{L} = I_{R} = I_{T}$ & $R_{L} = R_{R} =R_{T}$ & $C_{L} = C_{R} = C_{T}$ \\\hline  
 600 nA & 160 $\Omega$ & 5 fF to 500 fF \\ 
 \hline
\end{tabular}
\end{center}
\vspace{5mm}
Those values correspond to a quality factor ranging from 0.48 to 4.8.

\subsection{Three terminal Josephson junction analogue and its characterization}

We show on Figure S3 a more complete schematic of the circuit. The voltage controlled oscillators that it includes are home-made and described in greater details in Ref. \cite{Amet2021}. 

\begin{figure*}
    \centering
    \includegraphics[width=\textwidth]{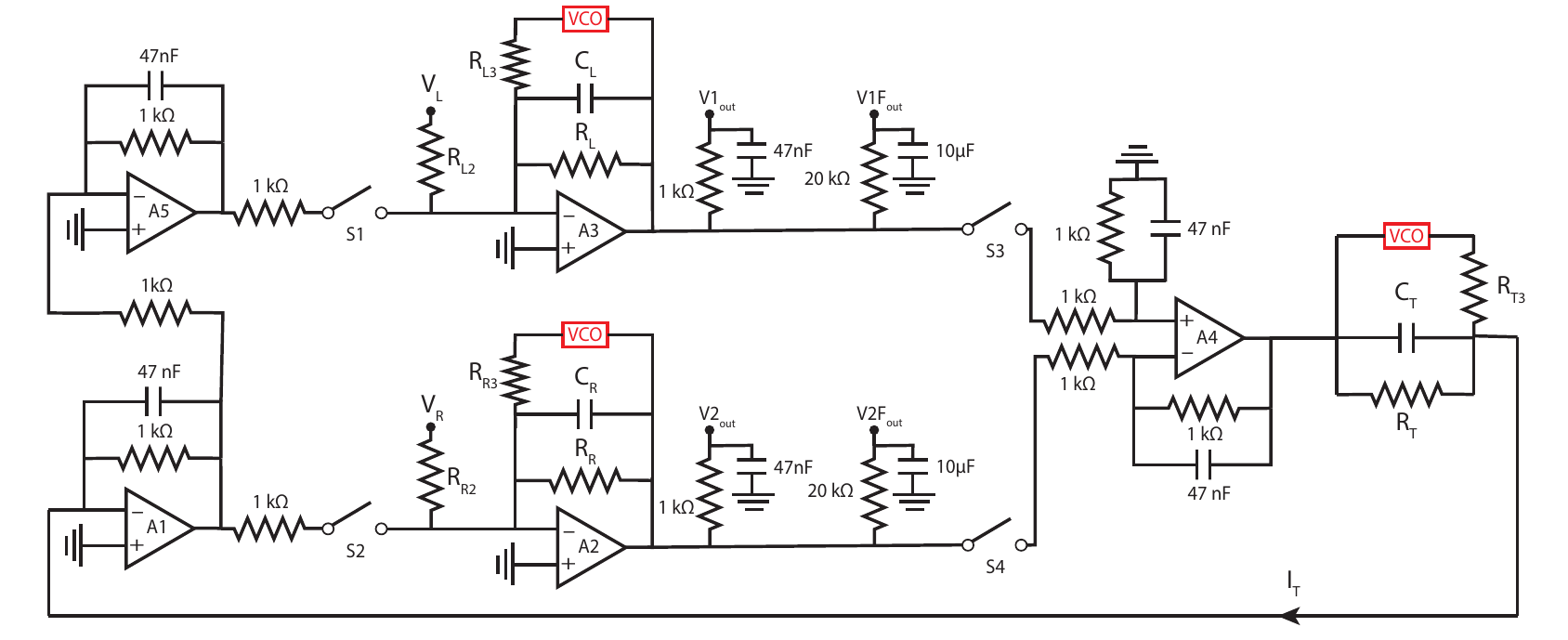}
 \caption {A more detailed diagram of the analog circuit. The detailed structure of the voltage controlled oscillators (in red) is not shown here.}
    \label{fig:FigSmjj}
\end{figure*} 

We define the output voltages of amplifiers A3 and A2 as $\frac{\dot{\phi}_{L}}{2\pi k}$ and $\frac{\dot{\phi}_{R}}{2\pi k}$, where k is the voltage to frequency gain of the VCO. With those notations, the voltages at the outputs of the oscillators in the feedback loops of amplifiers A3 and A2 are respectively $\alpha\sin(\phi_{L})$ and $\alpha\sin(\phi_{R})$, where $\alpha$ is the voltage amplitude of the output of the oscillators.

We call $I_{T}$ the current in the bottom branch of the circuit before the inverting terminal of A1. A1 is setup as a current to voltage converter and A5 as a unity gain inverting amplifier. Both amplifiers include shunting capacitors of 47nF to suppress high frequency noise, but those are irrelevant to the dynamics of the junctions which occurs on much longer timescales. We therefore ignore them for simplicity in the following derivation. This allows us to supply currents $I_{T}$ at the inverting input of A2, and $-I_{T}$ at the inverting input of A3. 

Assuming that the amplifiers A2 and A3 have a vanishing current input, we apply Kirchhoff rules at their inverting input and find:
\begin{equation}
    -I_{T}+\frac{V_{L}}{R_{L2}}+\frac{\alpha}{R_{L3}}\sin(\phi_{L})+\frac{\dot{\phi_{L}}}{2\pi k R_{L}}+C_{L}\frac{\ddot{\phi_{L}}}{2\pi k}=0
\end{equation}

\begin{equation}
    I_{T}+\frac{V_{R}}{R_{R2}}+\frac{\alpha}{R_{R3}}\sin(\phi_{R})+\frac{\dot{\phi_{R}}}{2\pi k R_{R}}+C_{R}\frac{\ddot{\phi_{R}}}{2\pi k}=0
\end{equation}

The amplifier A4 is setup as a differential amplifier, it thus generates a voltage $(\dot{\phi_{L}}-\dot{\phi_{R}})/(2\pi k)$, which is then fed to the subcircuit emulating the transverse junction. The 47nF capacitors are again suppressing high frequency noise but do not alter the dynamics of the junction. We find that:
\begin{equation}
I_{T}=\frac{\alpha}{R_{T3}}\sin(\phi_{L}-\phi_{R})+\frac{\dot{\phi_{L}}-\dot{\phi_{R}}}{2\pi k R_{T}}+C_{T}\frac{\ddot{\phi_{L}}-\ddot{\phi_{R}}}{2\pi k}
\end{equation}

Using the same matrix notation as before, we finally get:
\begin{align}
    \frac{1}{2\pi k}\mathcal{C} \ddot\Phi +\frac{1}{2\pi k}\mathcal{G} \dot\Phi+I_{c}(\Phi)=I
\end{align}

Where we defined $I_{L}=\frac{\alpha}{R_{L3}}$, $I_{T}=\frac{\alpha}{R_{T3}}$, and $I_{R}=\frac{\alpha}{R_{R3}}$, and:  

\begin{align*}
I_{c}(\Phi)&=\begin{pmatrix}I_{L}\sin(\phi_{L})+I_{T}\sin(\phi_{L}-\phi_{R})\\
    I_{R}\sin(\phi_{R})+I_{T}\sin(\phi_{R}-\phi_{L})
    \end{pmatrix}
\end{align*}

\begin{align*}
I &=\begin{pmatrix}-V_{L}/R_{L2}\\
-V_{R}/R_{R2}
    \end{pmatrix}
\end{align*}

We thus recover the same system of differential equations as for a three terminal shunted Josephson junction network, where the constant $\hbar/2e$ was replaced by $1/(2\pi k)$

\subsection{Analytical solution}

As discussed earlier, the time evolution of the phase is determined by the following matrix equation:

\begin{equation}
 \ddot\Phi +\mathcal{C}^{-1}\mathcal{G}\dot\Phi + \frac{2e}{\hbar}\mathcal{C}^{-1}I_c(\Phi)=\frac{2e}{\hbar}\mathcal{C}^{-1}I
\end{equation}

For simplicity we assume that all capacitances are equal to $C$, all conductances are equal to $G$, and all critical currents are equal to $I_{c}$. The matrices involved in equation (8) can thus be rewritten as follows::

\begin{align*}
I_{c}(\Phi)&=I_{c}\begin{pmatrix}\sin(\phi_{L})+\sin(\phi_{L}-\phi_{R})\\
    \sin(\phi_{R})+\sin(\phi_{R}-\phi_{L})
    \end{pmatrix}
    \\
\mathcal{C}&=C\begin{pmatrix}2 & -1\\-1& 2\end{pmatrix}, \mathcal{G}=G\begin{pmatrix}2 & -1\\
    -1& 2
    \end{pmatrix}, \mathcal{C}^{-1}=\frac{1}{3C}\begin{pmatrix}2 & 1\\1& 2\end{pmatrix}
\end{align*}

Equation (8) can be rewritten as:

\begin{align*}
 \begin{pmatrix} \ddot{\phi}_{L} \\ \ddot{\phi}_{R} \end{pmatrix} &+\frac{1}{RC}\begin{pmatrix} \dot{\phi}_{L} \\ \dot{\phi}_{R} \end{pmatrix} 
 \\
 &+ \frac{2e}{\hbar}\frac{I_{c}}{3C}\begin{pmatrix}2 & 1\\1& 2\end{pmatrix}\begin{pmatrix}\sin(\phi_{L})+\sin(\phi_{L}-\phi_{R})\\
    \sin(\phi_{R})+\sin(\phi_{R}-\phi_{L})
    \end{pmatrix}
    \\
    &=\frac{2e}{\hbar}\frac{1}{3C}\begin{pmatrix}2 & 1\\1& 2\end{pmatrix}I
\end{align*}

 Now we use the change of variables: $\eta=\frac{\phi_{L}-\phi_{R}}{2}$ and $\epsilon=\frac{\phi_{L}+\phi_{R}}{2}$. 
 
 We have:
 \begin{equation}
      \begin{pmatrix}\phi_{L} \\ \phi_{R} \end{pmatrix}=\begin{pmatrix}1 & 1\\-1& 1\end{pmatrix} \begin{pmatrix}\eta \\ \epsilon \end{pmatrix}
 \end{equation}
 
  So:
 
 \begin{align*}
 \begin{pmatrix} \ddot{\eta} \\ \ddot{\epsilon} \end{pmatrix} &+\frac{1}{RC}\begin{pmatrix} \dot{\eta} \\ \dot{\epsilon} \end{pmatrix}\\
 &+ \frac{e}{\hbar}\frac{I_{c}}{3C}\begin{pmatrix}1 & -1\\1& 1\end{pmatrix}\begin{pmatrix}2 & 1\\1& 2\end{pmatrix}\begin{pmatrix}\sin(\eta+\epsilon)+\sin(2\eta)\\
    \sin(\epsilon-\eta)-\sin(2\eta)
    \end{pmatrix}
    \\
    &=\frac{e}{\hbar}\frac{1}{3C}\begin{pmatrix}1 & -1\\1& 1\end{pmatrix}\begin{pmatrix}2 & 1\\1& 2\end{pmatrix}I
\end{align*}
 
 This becomes:
 
  \begin{align*}
 \begin{pmatrix} \ddot{\eta} \\ \ddot{\epsilon} \end{pmatrix} &+\frac{1}{RC}\begin{pmatrix} \dot{\eta} \\ \dot{\epsilon} \end{pmatrix} \\
 &+ \frac{e}{\hbar}\frac{I_{c}}{3C}\begin{pmatrix}\sin(\eta+\epsilon)-\sin(\epsilon-\eta)+2\sin(2\eta)\\
    3\sin(\epsilon-\eta)+3\sin(\eta+\epsilon)
    \end{pmatrix}\\
    &=\frac{e}{\hbar}\frac{1}{C}\begin{pmatrix}\frac{I_{L}-I_{R}}{3}\\ I_{L}+I_{R}
    \end{pmatrix}
\end{align*}

Using trigonometric identities we finally get:

\begin{align*}
 \begin{pmatrix} \ddot{\eta} \\ \ddot{\epsilon} \end{pmatrix} &+\frac{1}{RC}\begin{pmatrix} \dot{\eta} \\ \dot{\epsilon} \end{pmatrix} \\
 &+ \frac{e}{\hbar}\frac{I_{c}}{3C}\begin{pmatrix}2\sin(\eta)\cos(\epsilon)+2\sin(2\eta)\\
    6\sin(\epsilon)\cos(\eta)
    \end{pmatrix}\\
    &=\frac{e}{\hbar}\frac{1}{C}\begin{pmatrix}\frac{I_{L}-I_{R}}{3}\\ I_{L}+I_{R}\end{pmatrix}
\end{align*}

The equation for $\epsilon$ becomes:

  \begin{equation}
\ddot{\epsilon}+\frac{1}{RC}\dot{\epsilon}+\frac{2eI_{c}}{\hbar C}\cos(\eta)\sin(\epsilon)=\frac{e}{\hbar C}(I_{L}+I_{R})
\end{equation}

On the quartet resonance we have $I_{L}+I_{R}=0$ given the symmetry of the system. We define $I_{+}=\frac{e}{\hbar C}(I_{L}+I_{R})$ the deviation perpendicular to the quartet resonance, $\omega_{0}=\sqrt{\frac{2eI_{c}}{\hbar C}}$, $Q=\omega_{0}RC$. With these notations:

\begin{equation}
\ddot{\epsilon}+\frac{\omega_{0}}{Q}\dot{\epsilon}+\omega_{0}^{2}\cos(\eta)\sin(\epsilon)=I_{+}
\end{equation}

\begin{figure}
    \centering
    \includegraphics[width=\columnwidth]{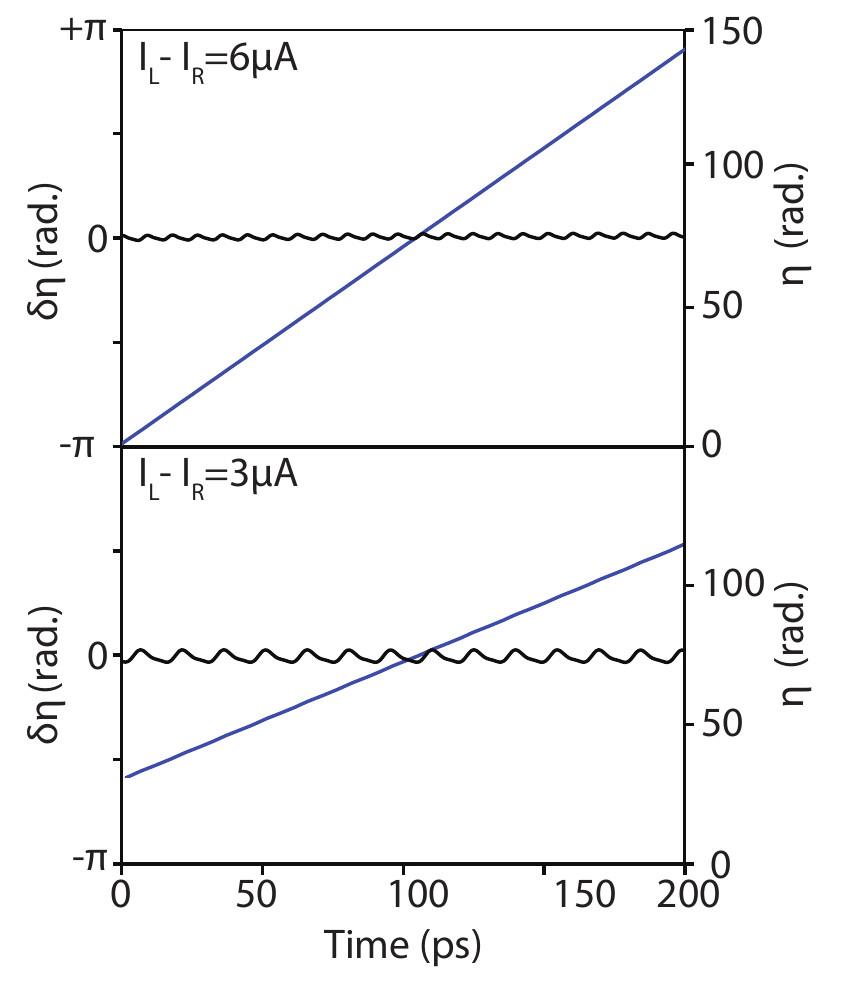}
    \caption {a) Simulated time evolution of $\eta$(t) plotted over 200 ps when $I_{L}-I_{R}= 6 ~\mu$A (blue). Difference $\delta\eta$ between $\eta$ and its linear fit (black). Simulation parameters are identical to those used in Figure 5. b) Same curves for a smaller bias asymmetry $I_{L}-I_{R}=3 ~\mu$A.}
    \label{fig:Figsim}
\end{figure}

We want to show that if $I_{+}$ is small, then $\langle \dot{\epsilon}\rangle$ remains zero. If that is the case that means the differential resistance at the quartet resonance is zero and the proof is complete. Recall that $\eta=\frac{\phi_{L}-\phi_{R}}{2}$. Along the quartet line, $\epsilon$ is small and $\eta$ varies rapidly since it is close to $\eta\approx \phi_{L}$. In fact $\eta \approx \omega t$ with $\omega\equiv\frac{e(V_{L}-V_{R})}{\hbar}$, up to an oscillating term which is negligible for high enough bias and Q factor. Two examples of the error which is made with this approximation are shown on Figure 9. Equation (23) becomes:

\begin{equation}
\ddot{\epsilon}+\frac{\omega_{0}}{Q}\dot{\epsilon}+\omega_{0}^{2}\cos(\omega t)\sin(\epsilon)=I_{+}
\end{equation}

This equation is the same as Kapitza's pendulum problem, describing the angle of an inverted rigid pendulum with an oscillating base. The only difference is the absence of a gravity term. 

Although the formalism to solve (12) has been extensively discussed elsewhere, we reproduce the solution here for completeness in the undamped case. We look for a solution as $\epsilon\equiv \epsilon_{S}(t)+A\cos(\omega t)+B\sin(\omega t)$, where $\epsilon_{S}$, A and B vary over much longer timescales than $2\pi/\omega$. A and B are also assumed to be small. We use this Ansatz in equation (12) and impose for (12) to be verified independently by slow-moving terms and $\cos(\omega t)$ terms and $\sin(\omega t)$ terms:

\begin{align*}
    \ddot{\epsilon}_{S}+\frac{A\omega_{0}^2}{2}\cos(\epsilon_{S})=I_{+}
    \\
    \ddot{A}-\omega^{2}A+2\dot{B}\omega+\omega_{0}^{2}\sin(\epsilon_{S})=0
    \\
    -2\dot{A}\omega+\ddot{B}-B\omega^{2}=0
\end{align*}

Keeping only highest order terms in $\omega$ we get:

\begin{align*}
    B&=0
    \\
    A&=\frac{\omega_{0}^{2}\sin(\epsilon_{S})}{\omega^{2}}
\end{align*}

The equation for $\epsilon_{S}$ becomes:

\begin{equation}
    \ddot{\epsilon_{S}}+\frac{\omega_{0}^4}{4\omega^{2}}\sin(2\epsilon_{S})= I_{+}
\end{equation}

When $I_{+}=0$, $\epsilon_{S}(t)$ tends to oscillate around two possible equilibria at both 0 and $\pi$, which implies that $\phi_{L}\equiv-\phi_{R} \mod 2\pi$. Equation (13) clearly has an equilibrium as long as $I_{+}$ is sufficiently small. This implies that $\langle \dot{\epsilon}\rangle=0$ even for small nonzero values of $I_{+}$. The differential resistance along the quartet line is therefore zero, which explains the resonance within this approximation. 

Finally, we note that equation 13 only has an equilibrium for small values of $ I_{+}$. Specifically:

\begin{equation}
    \lvert I_{+} \rvert <\frac{\omega_{0}^4}{4\omega^{2}}
\end{equation}

This translates to:
\begin{equation}
   \lvert I_{+}\rvert <\frac{\hbar I_{C}^{2}}{4eC(V_{L}-V_{R})^{2}}
\end{equation}

While other mechanisms might also be at play, for example self-heating, this trend alone is enough to explain the decrease of the quartet supercurrent at high bias. We see that the switching current is inversely proportional to the capacitance, which is seen in Figure 5c.

\bibliography{apssamp}

\end{document}